\documentclass[prd,aps,superscriptaddress,floatfix,nofootinbib,eqsecnum]{revtex4-2}

\pdfoutput=1

\usepackage{amsmath}
\usepackage{amsfonts}
\usepackage{amsmath}
\usepackage{amssymb}
\usepackage{bm}
\usepackage{dcolumn}
\usepackage{graphicx}
\usepackage[latin1]{inputenc}
\usepackage{latexsym}
\usepackage{rotating}
\usepackage{hyperref}
\usepackage{subfigure}
\usepackage{color}
\usepackage{changes}
\usepackage{verbatim}
\usepackage{comment}
\usepackage{empheq}
\usepackage{csquotes}
\usepackage{physics}
\usepackage{float}
\usepackage{amsmath,latexsym}
\usepackage{amsmath}
\usepackage{amssymb}
\usepackage{pifont}
\usepackage{soul}
\usepackage{booktabs}
\usepackage{tabularx} 
\begin{document}

\title{Deformed algebraic structure of angular momenta:  GUP perspective
}

\author{Gaurav Bhandari}\email{bhandarigaurav1408@gmail.com}\affiliation{Department of Physics, Lovely Professional University, Phagwara, Punjab, 144411, India}

\author{S. D. Pathak}\email{shankar.23439@lpu.co.in}\affiliation{Department of Physics, Lovely Professional University, Phagwara, Punjab, 144411, India}

\author{Manabendra Sharma}\email{sharma.man@mahidol.ac.th}\affiliation{Centre for Theoretical Physics and Natural Philosophy, Nakhonsawan Studiorum for Advanced Studies, Mahidol University, Nakhonsawan, 60130, Thailand}

\author{Anzhong Wang}\email{Anzhong$_$Wang@baylor.edu}
\affiliation{GCAP-CASPER, Department of Physics, Baylor University,
Waco, Texas 76798-7316, USA}

\begin{abstract}
The prediction of a minimal length scale by various quantum gravity candidates (such as string/M theory, Doubly Special Relativity, Loop Quantum Gravity and others) have suggested modification of Heisenberg Uncertainty Principle (HUP), resulting in the Generalized Uncertainty Principle (GUP).

In this  short review, we investigate the origins of the GUP and examine higher-order models, focusing on the linear plus quadratic form of the GUP. We extend the concept of minimal length to minimal angular resolution, which plays a crucial role in modifying angular momentum and its associated algebra. A comparison is made between the standard angular momentum commutator algebra and that modified by the GUP. Finally, we review its application in the hydrogen atom spectra and and discuss future endeavors. 
\end{abstract}

\maketitle


\tableofcontents
\section{Introduction}
General relativity (GR) is a well-established theory that has successfully explained many astrophysical phenomena and offers viable mathematical models of cosmology to determine the dynamics of the universe. At large scales gravity became predominant and thus GR serves as a viable framework at the classical level. In fact,  the explanation of perihelion precession of mercury\cite{rana1987investigation}, deflection of light when passing through massive bodies\cite{genov2009mimicking}, and gravitational redshift of light \cite{wojtak2011gravitational} are few epitomes of its grand success in last hundred years.  

Despite its successes, particularly in describing objects like black holes, the theory appears incomplete when addressing the singularities of these objects \cite{cite1,cite2,cite3,cite4,cite5,63,64}. GR also struggles to explain the spacetime singularity at the beginning of the universe. While the universe appears homogeneous and isotropic at large scales, minute fluctuations of order $10^{-5}$ are observed in CMB. These fluctuations necessitate the use of perturbative theory, where the zeroth order is the homogeneous and isotropic universe given by Friedmann-Lema\^itre-Robertson-Walker (FLRW) metric. In the perturbative approach, the background metric is treated classically, while the first-order correction term is quantized in linearized gravity theory \cite{guth1982fluctuations,linde1982new,bardeen1983spontaneous,ashtekar2005quantum,sing1,ashtekar2009loop}.

This compels the incorporation of two fundamentally different, and yet incompatible, frameworks, quantum theory and general relativity (GR) into a coherent setting. There have been numerous attempts to develop a complete quantum theory of gravity, like String/M Theory (ST) \cite{veneziano1986stringy,witten1996reflections,scardigli1999generalized,gross1988string,amati1989can,yoneya1989interpretation,s3}, Loop Quantum Gravity (LQG) \cite{rovelli1998strings,sharma2019background,garay1995quantum,as1,ha1,sh1}, and Doubly Special Relativity (DSR) \cite{double1, gh1}, each with its advantages and issues. One of the commonest  issues amongst the quantum gravity candidates is the recovery of the lower energy regime. While these theories have made significant progress, they lack experimental evidence and yet to undergo rigorous phenomenological testing. Even if supersymmetry is observed in the Large Hadron Collider (LHC), it would only confirm the existence of an essential ingredient of string theory and would not constitute definitive evidence in favor of the theory itself \cite{68,69,70,71,72}. At Planck scales, the search for a theory of quantum gravity becomes crucial, as it may address the limitations of GR.
With the existence of so many approaches to QG, it is important to try to extract testable predictions. Recently, many attempts have been made in this direction. Although some do compute quantum gravity effects, the minuscule scale of the Planck length (and the immense scale of the Planck energy) often renders these effects negligible.

In this series of endeavors, one of the simplest and potentially testable approaches to quantum gravity (QG) is the Generalized Uncertainty Principle (GUP), which incorporates the minimal length prediction of many other QG theories in the well-known Heisenberg Uncertainty Principle (HUP). According to HUP, two canonically conjugate variables position and momentum, we can achieve arbitrary precision in position measurement by giving a maximal amount of energy or when momentum is unknown. However, many QG theories predict that this arbitrary precision is not valid below the Planck length, where the classical notion of space-time no longer applies \cite{54,55,56,57,58,59,60,61,62,65,66,67,100,101,102}. On considering this fact HUP can be modified by incorporating additional terms on the right-hand side of the Heisenberg uncertainty relation giving rise to GUP. Accounting for quantum fluctuations in spacetime leads to the theory of the GUP, where these fluctuations in the geometry of spacetime increase the uncertainty in measuring both position and momentum.  Since the GUP modifications exist in canonically conjugate variables, most quantum gravity (QG) theories predict momentum-dependent changes to the position-momentum commutation relation. These modifications in the existing commutator brackets and the HUP affect well-known aspects of quantum mechanics, as demonstrated by \cite{30,31,32,32a}.
This leads to alterations in the HUP and suggests the existence of a minimum measurable length near the Planck scale \cite{34,35,36}. 

Studies of the GUP have revealed changes in several aspects of Quantum Mechanics (QM). For instance, the Hamiltonian describing a minimal coupling with an electromagnetic field is expected to undergo modifications, as demonstrated in \cite{30}. This has been used to show changes in the Landau levels \cite{30}. Additionally, the GUP is found to affect phenomena such as the Lamb shift, potential steps and barriers, which are relevant in quantum configurations like Scanning Tunnelling Microscopes \cite{31,32}, as well as the case of a particle in a box, where the box's length becomes quantized \cite{31,32,32a}. Moreover, the energy levels of a simple harmonic oscillator are also altered \cite{31}. It has been suggested that the GUP could be detected through its effects on quantum optical systems \cite{33,G8}.

In this paper, we discuss GUP corrections to a significant theoretical and experimental area of quantum mechanics mainly in the line of he seminal work in \cite{53}. Specifically, we examine how GUP leads to modifications in the angular momentum algebra. This article is organized as follows. In Sec. \ref{sec1}, we introduce the origin of GUP and explain how to find the modified canonical variables. In Sec. \ref{sec2}, we recap the existing angular momentum algebra under the Heisenberg Uncertainty Principle (HUP). In Sec. \ref{sec3}, we discuss the modifications arising from GUP corrections. Finally, in Sec. \ref{sec4}, we extend the application of the modified angular momentum algebra to the hydrogen atom and observe the resulting modifications to the energy levels.

\section{Origin of Generalized Uncertainty Principle}\label{sec1}
Many theories of QG predict the existence of a minimal length. In string theory, it is conjectured that strings do not interact at distances smaller than their size, which is determined by their tension \cite{37,38}. Essentially, particles are considered as vibration of strings with fundamental lengths close to the Planck length. Various \textit{gedanken} experiments involving black holes also suggest the existence of a minimal length. LQG inherently includes a minimal length through its discrete area and volume elements. Similarly, the DSR theory, a modification of the Special Theory of Relativity, incorporates minimal length scales and maximum measurable momentum.

The GUP represents an innovative approach in the quest for a quantum gravity theory. It involves modifications to the well-known HUP. The HUP, in its traditional form, does not accommodate the concept of a minimal length, as it allows for arbitrary values of uncertainty in position. The GUP addresses this by proposing modifications to the HUP, incorporating higher-order corrections inspired by various QG theories.

In 1995 Kempf, Mangano and Mann (KMM) first proposed the generalized uncertainty relation that results in minimal observable length \cite{Z2, Z3,Z4} as

\begin{equation}\label{kmm}
\Delta X \Delta P \geq \frac{\hbar}{2}(1+\beta (\Delta P)^2 + \Omega), 
\end{equation}
where $\beta$ is the GUP parameter and $\Omega = \beta \langle P \rangle ^2$ which is a positive constant depending on the expectation value of the momentum operator. We also define $\beta= \beta_0/ (M_{Pl} c)^2$ where $M_{Pl}$ is the Planck mass and $\beta_0$ is of the order of unity.
One can easily obtained the minimal observable length as $(\Delta X)^{KMM}_{min} = \hbar \sqrt{\beta}$. The above uncertainty relation is obtained in one dimension by the deformed commutator bracket 
\begin{equation} \label{KMM}
[X, P] = i \hbar (1+\beta P^2).
\end{equation}
As shown by KMM in their paper \cite{ Z2, Z3,Z4}, the following operator equations, for $X$ and $P$ can be immediately obtained from commutator algebra Eq.(\ref{KMM}):
\begin{align}
P\phi(p) &= p \phi (p),\\
X\phi(p) &= i \hbar(1+\beta p^2)\partial_p \phi(p),
    \end{align}
where \( X \) and \( P \) be symmetric operators defined on a dense domain \( S_\infty \) (Schwartz space), with respect to the scalar product

\begin{equation}
\langle \psi | \phi \rangle = \int_{-\infty}^{\infty} \frac{dp}{1 + \beta p^2} \psi^*(p) \phi(p),
\end{equation}
where the relation \( \int_{-\infty}^{\infty} \frac{dp}{1 + \beta p^2} |p \rangle \langle p | = 1 \) holds, and \( \langle p | p' \rangle = (1 + \beta p^2) \delta(p - p') \). Given this scalar product definition, the commutation relation in Eq. (\ref{KMM}) is exactly satisfied. Specifically, the generalized version of the HUP shown in Eq.(\ref{kmm}) is considered to represent only the leading term in an expansion involving the small parameter $\sqrt{\beta} \Delta P$ of a higher-order GUP. Several efforts have been made to develop a complete version of the KMM GUP valid at all energy scales. 

Based on the field theory on non-anticommutative superspace, Nouicer first proposed a more generalized form of GUP that broadens the validity domain of the KMM GUP. The deformed commutator bracket \cite{42} is 
\begin{equation}
[X, P] = i \hbar 
 \exp(\beta P^2),
\end{equation}
and from the Schr\"{o}dinger-Robertson uncertainty relation between two operators gives
\begin{equation}
\Delta X \Delta P \geq \frac{\hbar}{2} \exp(\beta \Delta P^2).
\end{equation}
This algebra can be satisfied by the position and momentum operators in momentum space representation as
\begin{align}
P\phi(p) &= p \phi (p),\\
X\phi(p) &= i \hbar\exp(\beta p^2)\partial_p \phi(p).
\end{align}
Now using the symmetricity condition of position operator implies the modified scalar product as
\begin{align}
\langle \psi | \phi \rangle &= \int_{-\infty}^{\infty} \frac{dp} \exp(\beta p^2) \psi^*(p) \phi(p),\\
\langle p | p' \rangle &= \exp(\beta p^2) \delta(p-p').
\end{align}
This GUP gives the minimum uncertainty in the position as $(\Delta X)^{Nouicer}_{min}= \sqrt{\frac{e}{2}}\hbar \sqrt{\beta}$.
Later this GUP was again modified to incorporate minimum uncertainty in position as well as maximum momentum as proposed by Pedram \cite{43} as 
\begin{equation}\label{ped}
[X,P]= \frac{i \hbar}{1-\beta P^2},
\end{equation}
this commutator relation agrees with KMM's and Nouicer's GUP to the leading orders and contains singularity at $p^2=1/\beta$. This indicates that the momentum of the particle cannot exceed \(\frac{1}{\sqrt{\beta}} \approx \frac{1}{\alpha}\). And, this gives the uncertainty relation as 
\begin{equation}
\Delta X \Delta P \geq \frac{\hbar/2}{1- \beta (\Delta P)^2}.
\end{equation}

The position and momentum operators in the momentum representation for Pedram GUP Eq.(\ref{ped}) are
\begin{align}
P\phi(p) &= p \phi (p),\\
X\phi(p) &= \frac{i \hbar}{1- \beta p^2}\partial_p \phi(p),    
\end{align}
 and using the symmetricity condition the modified scalar product as
 \begin{align}
 \langle \psi | \phi \rangle &= \int_{-1/\sqrt{\beta}}^{1/\sqrt{\beta}} dp \exp(-\beta p^2) \psi^*(p) \phi(p),\\
\langle p | p' \rangle &= \frac{\delta(p-p')}{1-\beta p^2}.    
 \end{align}
 A concise  comparison between the different GUPs can be better given  in tabular form \ref{GUPs}. 
\begin{table}[h]
\centering
\caption{Comparison of different GUP models.}
\begin{tabular}{cccc}
\toprule
\textbf{GUP Model} & \textbf{Deformed Commutator} & \textbf{Minimal Length Uncertainty} & \textbf{Maximal Observable Momentum} \\ 
\addlinespace
\midrule
\addlinespace
HUP & $[X, P] = i\hbar$ & \ding{55} & \ding{55} \\ 
\addlinespace
KMM & $[X, P] = i\hbar(1 + \beta P^2)$ & \checkmark & \ding{55} \\ 
\addlinespace
Nouicer & $[X, P] = i\hbar \exp(\beta P^2)$ & \checkmark & \ding{55} \\ 
\addlinespace
Pedram  & $[X, P] = \frac{i\hbar}{1 - \beta P^2}$ & \checkmark & \checkmark \\ 
 
\addlinespace
\bottomrule
\end{tabular}

\label{GUPs}
\end{table}

We also note that the minimal length $(\Delta X )_{min}^{KMM}< (\Delta X )_{min}^{Nouicer}< (\Delta X )_{min}^{Pedram}$ shown in table \ref{gup_comparison}.

\begin{table}[ht]
    \centering
    \caption{Comparison of minimal length uncertainties and maximal momenta in three GUP frameworks.}
    \renewcommand{\arraystretch}{1.5} 
    \setlength{\tabcolsep}{12pt} 
    \begin{tabular}{|>{\centering\arraybackslash}p{4cm}|>{\centering\arraybackslash}p{4cm}|>{\centering\arraybackslash}p{4cm}|}
        \hline
        
        \textbf{GUP Framework} & \textbf{$(\Delta X )_{min}$} & \textbf{$P_{max}$} \\ 
        \hline
        KMM & $\hbar \sqrt{\beta}$ & - \\ 
        \hline
        Nouicer & $\hbar \sqrt{\beta} \sqrt{\frac{e}{2}}$ & - \\ 
        \hline
        Pedram & $\frac{3\sqrt{3}}{4} \hbar \sqrt{\beta}$ & $\frac{1}{\sqrt{\beta}}$ \\ 
        \hline
    \end{tabular}
    \label{gup_comparison}
\end{table}

In 2011 Ali, Das, and Vagenas proposed \cite{44} a generalized model incorporating quadratic terms of momenta as dictated by ST and various \textit{Gedanken experimente} in black hole physics, and linear terms suggested by DSR as in \cite{73,74,75,76,77,78,79}. The GUP deformed commutator bracket is expressed as:
\begin{equation}\label{das}
[q_i,p_j]=i\hbar\Bigg\{\delta_{ij}- \alpha \left(p\delta_{ij}+\frac{p_ip_j}{p}\right) + \alpha^2[ p^2 \delta_{ij}+3p_ip_j]\Bigg\},
 \end{equation}
 where $\alpha= \alpha_0/M_{Pl}c= \alpha_0 l_{Pl}/\hbar$, $M_{Pl}$ = Planck mass, $l_{Pl} \approx 10^{-35} m = \text{Planck length}$, and $M_{Pl}c^2  \approx 10^{19} GeV = \text{Planck energy}$. This form of GUP gives:
 \begin{align}
   \Delta x \geq (\Delta x)_{\text{min}} &\approx \alpha_0 l_{\text{Pl}} \\
   \Delta p \leq (\Delta p)_{\text{max}} &\approx \frac{M_{\text{Pl}}c}{\alpha_0}.
\end{align}

\begin{figure}[!h]
\begin{center}
  \begin{minipage}{0.3\textwidth}
    \centering
    \includegraphics[width=\textwidth]{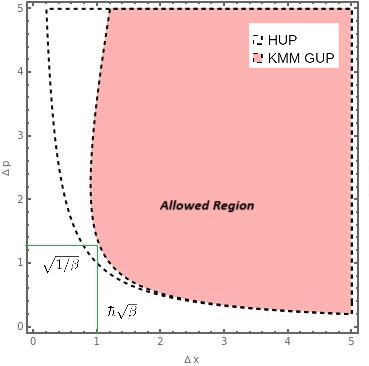}
  \end{minipage}
  \begin{minipage}{0.3\textwidth}
    \centering
    \includegraphics[width=\textwidth]{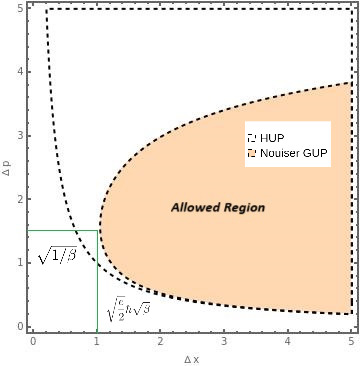}
  \end{minipage}
  \begin{minipage}{0.3\textwidth}
    \centering
    \includegraphics[width=\textwidth]{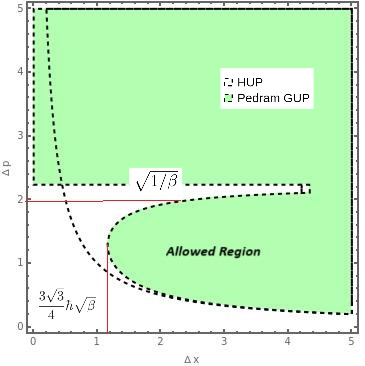}
  \end{minipage}
\end{center}
\caption{In this plot, Figure (a) on the left shows a comparison between HUP and KMM GUP, where the minimal length is given by $\hbar \sqrt{\beta}$. Figure (b) represents the comparison between Nouicer GUP and HUP, which again only includes minimal length but no maximal observable momentum. The final figure compares Pedram GUP with HUP, and we observe that in this GUP, both minimal length and maximal observable momentum are present.
}  
\label{plot}
\end{figure}
Now we can express the modified momentum $p_i$ in terms of low energy momentum term $p_{0i}$, since Eq.(\ref{das}) contain quadratic momentum $p_i$, the latter must contain at most cubic terms of $p_{0i}$. So the most general form consistent with the structure is:
\begin{equation}\label{p1}
p_j =p_{0j}+ap_0 p_{0j}+bp_0^2p_{0j},
\end{equation}
where $\alpha \sim a$, $b \sim a^2$. Now using Eq.(\ref{das}) 
\begin{equation}
[x_i,p_j]=[x_i,p_{0j}]+a([x_i,p_0]p_{0j}+p_0[x_i,p_0j])+b([x_i,p_0]p_0p_{oj}+p_0[x_i,p_0]p_{0j}+p_0^2[x_i,p_{0j}]),
\end{equation}
the more general form of the commutator bracket of Eq.(\ref{das}) incorporating linear and quadratic term is 
\begin{equation}\label{111}
[x_i,p_j]=i\hbar(\delta_{ij}+\delta_{ij}\alpha_1 p +\alpha_2 \frac{p_i p_j}{p}+ \beta_1 \delta_{ij}p^2 +\beta_2 p_i p_j),
\end{equation}
using Eq.(\ref{111}) we obtain
\begin{align}
    [x_i, p^2] &= [x_i, p \cdot p] = [x_i, p] p + p [x_i, p]\label{112} \\
    &= [x_i, p_k \cdot p_k] = [x_i, p_k] p_k + p_k [x_i, p_k] \notag \\
    &= 2i\hbar p_i \left[1 + (\alpha_1 + \alpha_2) p \right] \quad \text{using Eq.(\ref{111}) to }\mathcal{O}(p), \label{113}
\end{align}
and on comparing Eq.(\ref{112}) and Eq.(\ref{113}) we get,
\begin{equation}
[x_i,p]=i \hbar (p_ip^{-1}+(\alpha_1+\alpha_2)p_i)
\end{equation}
and when $\alpha_i=0 $, we obtain
\begin{equation}\label{a}
[x_i,p]=i \hbar p_{0i}p_0^{-1},
\end{equation}
we know
\begin{equation}\label{b}
p_j =P_{0j}(1+ap_0)+\mathcal{O}(a^2) \simeq P_{0j}(1+ap) \quad \text{[from Eq.(\ref{p1})]}
\end{equation}
which gives
\begin{equation}\label{c}
p_0j\simeq \frac{p_j}{1+ap}\simeq (1-ap)p_j,
\end{equation}
\begin{align}\label{d}
p_0&=(p_{0j} p_{0j})^{\frac{1}{2}}\\
&=((1-ap)^2p_jp_j)^{\frac{1}{2}}=(1-ap)p,
\end{align}
we also obtain
\begin{equation}\label{e}
p_{0i}p_0^{-1}p_{0j}=(1-ap)p_i(1-ap)^{-1}p^{-1}(1-ap)p_j=(1-ap)p_ip_jp^{-1}.
\end{equation}
and, by assuming coordinates and momenta commutes among themselves, it follows the Jacobi identity:
\begin{equation}\label{jacobi}
[[x_i, x_j], p_k] + [[x_j, p_k], x_i] + [[p_k, x_i], x_j] = 0.
\end{equation}
Now using Eq.(\ref{jacobi}),we obtain
\begin{equation}\label{correct}
  [x_i,p_j]=i\hbar\delta_{ij} +i\hbar a(p\delta_{ij}+p_ip_jp^{-1})+i\hbar (2b-a^2)p_i p_j +i\hbar (b-a^2)p^2 \delta_{ij},  
\end{equation}
comparing Eq.(\ref{correct}) with Eq.(\ref{das}) gives $a=-\alpha$ and $b=2a^2$. We obtain the deformed momentum and position in terms of low energy-momentum terms as:
\begin{equation}
x_i = x_{0i}, \quad p_i= p_{0i}(1-\alpha p_0 +2 \alpha ^2 p_0^2).
\end{equation}

Since the idea of minimal observational length can extend to minimal angular resolution, we will choose the angular variable and its conjugate momentum (angular momentum). We expect modifications to the well-known angular momentum algebra as a consequence of GUP.

\section{Angular momentum algebra in HUP}\label{sec2}

In this section, we review the standard angular momentum commutator brackets using the HUP and then compare it with the results obtained when applying the GUP. Angular momentum in classical and quantum mechanics plays an important role in understanding many physical phenomenons and their mathematical descriptions like the trajectory of planets and planetary systems (Kepler's law), rotation of rigid bodies, the structure of atoms, and many more. In this section, we start with the classical definition of angular momentum. In classical mechanics, angular momentum measures the ``amount of rotation" likewise linear momentum which measures the ``amount of motion". In simple terms, for point particle, it is the cross product of position and momentum. Angular momentum also follows the law of conservation as motion is in the central field. The particle motion is restricted to the plane perpendicular to the angular momentum vector. 

The classical description of angular momentum is given as follows:
the angular momentum $\Vec{L}$ of a particle with mass 
$\textit{m}$
and linear momentum 
$\Vec{p}$
 located at position 
$\Vec{r}$ is given by the vector cross product:
 \begin{equation}
    \Vec{L}= \Vec{r} \cross \Vec{p}.
 \end{equation}
where the magnitude is given by $L= rp \sin{\theta}$ and $\theta$ is the angle between the vectors $\Vec{r}$ and $\Vec{p}$. The components in the cartesian coordinate are:
 \begin{equation}
 L_x=yp_z-zp_y, \quad L_y= zp_x-xp_z, \quad L_z=xp_y-yp_x . \label{components}\end{equation}
In quantum mechanics, physical quantities such as angular momentum are represented by operators acting on the wave function of a system, rather than continuous variables. The classical quantities of position and momentum are promoted to operators as we go into the quantum regime. The position operator $\hat{r}$
  and the momentum operator 
$\hat{p}$
  in one dimension are defined as:
\begin{equation}
r \equiv \hat{r} \equiv \hat{q}, \quad p \equiv \hat{p} \equiv -i\hbar\frac{d}{dq}.
\label{oper}\end{equation}
Angular momentum is defined through the generator of infinitesimal rotations and its algebraic properties. However, for orbital angular momentum, this definition in Eq. (\ref{components}) and the generator of rotation are equivalent. This is analogous to linear momentum, which is the generator of translations and, like angular momentum, does not give rise to a minimal uncertainty relation, as it follows the standard Heisenberg algebra. Angular momentum, as the generator of rotation, also does not allow for minimal angular resolution. Following this, the corresponding components of angular momentum in Cartesian coordinates are represented as follows:
\begin{equation}\label{5}
L_x= -i \hbar \left(y\frac{\partial}{\partial z}-z\frac{\partial}{\partial y}\right), \quad L_y= -i \hbar \left(z\frac{\partial}{\partial x}-x\frac{\partial}{\partial z}\right), \quad L_z= -i \hbar \left(x\frac{\partial}{\partial y}-y\frac{\partial}{\partial x}\right).
\end{equation}

Using the commutator relation between position and linear momentum ($[q_i,p_j]=i\hbar \delta_{ij}$), one can  easily obtain the ``fundamental commutation relations of angular momentum" which incorporate all the basic properties of rotation in three dimensions \cite{45,46,47,48,49,50}: 
\begin{equation}\label{e1}
[L_i,L_j] = i\hbar \epsilon_{ijk}L_k.
\end{equation}
with the Levi-Civita symbols 
\[
\epsilon_{ijk} = \left\{
\begin{array}{ll}
+1 & \text{if ijk is an even permutation of 123}   \\
-1 &  \text{if ijk is an even permutation of 123} \\
 0 & \text{otherwise}.
\end{array}
\right.
\]
To define Eq.(\ref{e1}), we used two main concepts: first, that $L_k$ is the generator of rotations about the $K^{\text{th}}$ axis, and second, that rotations about different axes do not commute. Additionally, one can verify that the Jacobi identity, an important property of the Lie algebra of angular momentum operators, follows from Eq.(\ref{e1}) is given as
\begin{equation}\label{e2}
[L_i,[L_j,L_k]]+[L_j,[L_k,L_i]]+[L_k,[L_i,L_j]]=0,
\end{equation}
Using Eq.(\ref{e2}), it is easy to show that we can define eigenstates of $L^2$  and $L_z$ simultaneously or in other words $L^2$ commutes with all of its components as:
\begin{equation}\label{e3}
  [L^2,L_j]=[L_i^2,L_j]+[L_j^2,L_j]+[L_k^2,L_j]=0.  
\end{equation}
In fact, consider that
\begin{equation}\label{w1}
[L_i,p_m]=\epsilon_{ijk}[q_j p_k,p_m]= \epsilon_{ijk}[q_j,p_m]p_k= i\hbar \epsilon_{ijk}p_k
\end{equation}
Using Eq.(\ref{w1}), we have
\begin{equation}\label{e4}
[L_i,p^2] = [L_i,p_i^2+p_j^2+p_k^2]=p_i[L_j,p_i]+[L_j,p_i]p_i+p_j[L_k,p_j]+[L_k,p_j]p_j =0
\end{equation}
and from this 
\begin{equation*}
[L_i,p^2]=0=2p[L_i,p]
\end{equation*}
that means for $p \neq 0$ gives
 \begin{equation}\label{e5}
    [L_i,p]=0.
 \end{equation}
 Similarly, commutator relation of  $L^2$ with $p$ and $p^2$ is as follows:
 \begin{align}
[L^2,p]&= L_i[L_i,p]+[L_i,p]L_i=0, \label{e6}\\  
[L^2,p^2]&=L_i[L_i,p^2]+[L_i,p^2]L_i=0.\label{e7} 
\end{align}
From the above two Eqs.(\ref{e6},\ref{e7}), we observe that the operators $L^2$, $L_z$, $p$ and $p^2$ commute, allowing us to construct simultaneous eigenstates of these quantities. These eigenstates are also eigenstates of the free particle Hamiltonian, which represents the system's energy in the absence of external forces or potentials. Since the angular momentum operator corresponds to an observable physical quantity, it is Hermitian, meaning it produces real eigenvalues when acting on its eigenstates. In the context of group theory, operators that commute with one another form an abelian group. However, because the components of angular momentum do not commute, the algebra they form is non-abelian.
\subsection{Ladder operators}
Quantum operators like angular momentum are not a convenient way to keep track of eigenvalues and eigenfunctions. For this we introduced Ladder operators as $L_+$ and $L_-$ on the eigenfunction $Y_{lm}$ where $l$ and $m$ are the quantum numbers.
The usual ladder operators are given as:
\begin{equation*}
L_+= L_x+ i L_y, \quad  L_-=L_x- i L_y,
\end{equation*}
with commutator relation with the z-component of angular momentum as:
\begin{equation}\label{lz}
[L_z,L_{\pm}]=[L_z,L_x] \pm i [L_z,L_y]= \hbar(i L_y \pm L_x)= \pm \hbar L_{\pm},
\end{equation}
The $L^2$ commutes with the ladder operators $L_{\pm}$ as
\begin{equation}\label{ls}
[L^2,L_{\pm}]= [L^2,L_x] \pm i [L^2, L_y] =0.
\end{equation}
We also have
\begin{equation}\label{1}
[L_+,L_-] = [L_x+iL_y,L_x-iL_y]= -2i[L_x,L_y]=2 \hbar L_z .
\end{equation}
The ladder operator also satisfies,
\begin{equation}\label{2}
L_{\pm}L_{\mp} = L^2-L_z^2 \pm \hbar L_z.
\end{equation}
We also know that ladder operators are not Hermitian since they do not correspond to any physical observable quantities; rather, they act to raise or lower the eigenstates. The eigenvalues of $L^2$ and component  $L_z$ for the given eigenstate are
\begin{equation}\label{3}
L^2  |nlm\rangle = \hbar ^2 l(l+1)|nlm\rangle, \quad L_z|nlm\rangle = \hbar m |nlm\rangle
\end{equation}
where $n$, $l$, and $m$ are quantum numbers that take on only integer values. 
Now using Eq.(\ref{lz}), we get
\begin{equation}\label{4}
L_zL_{\pm}  |nlm\rangle =(m \pm 1) \hbar L_{\pm}|nlm\rangle 
\end{equation}
which means that $L_{\pm}|nlm\rangle$ act as an eigenstate for $L_z$ with eigenvalue $(m\pm 1)\hbar$. The operators \(L_{\pm}\) are also referred to as raising and lowering operators, as they increase or decrease the eigenvalue of \(L_z\) by an amount of \(\hbar\). Now using Eq.(\ref{ls}) we also obtain
\begin{equation}\label{5}
L^2L_{\pm} |nlm\rangle = L_{\pm}L^2|nlm\rangle= l(l+1)\hbar ^2 L_{\pm}|nlm\rangle   \end{equation}
From the above equation, we observe that the ladder operators act on the eigenstates of $L^2$ without altering its eigenvalue, while they do change the eigenvalue associated with $L_z$. This point emphasizes that ladder operators affect the quantum number $m$ associated with the $L_z$ component but leave the total angular momentum quantum number $l$ unchanged. We can determine the eigenvalue for the ladder operators in a given state, where the ladder operator shifts the state either one level up or one level down:
\begin{equation}\label{6}
L_{\pm} |l,m\rangle = \gamma_{\pm} \hbar|l, m\pm 1\rangle
\end{equation}
where, $\gamma_{\pm} = \sqrt{l(l+1)-m(m\pm1)}$.

In this section, we discussed the standard results of angular momentum algebra in quantum mechanics. All of the aforementioned results are derived under the assumptions of the HUP without considering the implications of a minimal length or minimal angular resolution. In the next section, we will address the concept of minimal length and its effects on angular momentum algebra.
\section{ Modified Angular momentum algebra in the presence of GUP}\label{sec3}
In this section, we study how the effects of the GUP (GUP), where the modified commutator relation between position and momentum alters the standard quantum mechanics of angular momentum and its commutator relations. Specifically, we focus on a form of GUP where the coordinates remain unchanged, and only the momentum is modified \cite{51, 52,53}as follows:
\begin{equation}
q_i=q_{0,i}  \quad  p_i=p_{0,i}[1-\delta p_0 +(\epsilon +\delta ^2)p_0^2],\label{deformed}
\end{equation}
 with the commutator relation as:
 \begin{equation}\label{deform}
[q_i,p_j]=i\hbar\Bigg\{\delta_{ij}- \gamma \delta \left(p\delta_{ij}+\frac{p_ip_j}{p}\right) + \gamma ^2[\epsilon p^2 \delta_{ij}+(2\epsilon +\delta^2)p_ip_j]\Bigg\}
 \end{equation}
where $\gamma$ is related to the scales at which the quantum-gravitational effect became relevant, and defined as the inverse of Plank momentum
\begin{equation}
\gamma=\frac{\gamma_0}{M_{Pl} c} \quad  \text{with} \quad \gamma_0 \sim 1,
\end{equation}
the dimensionless parameters $\delta$ and $\epsilon$ are included to highlight the terms originating from the linear and quadratic contribution to GUP. For the one-dimensional case with $\delta=0$ and $\epsilon=1/3$, we easily recover the KMM GUP shown in Eq.(\ref{KMM}). This model of GUP does not imply a non-commutation relation because of $[x_i,x_j]=0=[p_i,p_j]$. The motivation for choosing this type of GUP, which includes multiple free parameters, lies in the flexibility it provides. By adjusting these parameters, we can explore and analyze various forms of the GUP. Although this GUP is not the higher-order version proposed by Pedram and Nouicer, we anticipate that a higher-order GUP would yield similar results, with the corrections becoming more precise at higher orders.


The GUP between position and momentum introduces a minimal measurable length. Similarly, if we apply GUP to angular momentum algebra, it leads to the concept of a minimal measurable angle. We expect GUP in the angular variables and their canonical angular momentum. 
To see this, we use the GUP-modified commutator relation of Eq.(\ref{e1}), which changes the commutator relation between the components of angular momnetum as

\begin{equation}\label{use}
\begin{split}
[L_i,L_j] &= \epsilon_{imn}\epsilon_{jrs}[q_mp_n,q_rp_s] \\
&= \epsilon_{imn}\epsilon_{jrs}\{q_m[p_n,q_r]p_s+q_r[q_m,p_s]p_n\} \\
&= i \hbar \epsilon_{imn}\epsilon_{jrs} \left\{q_rp_n\left[\delta_{ms}- \gamma \delta \left(p\delta_{ms}+\frac{p_mp_s}{p}\right) + \gamma ^2[\epsilon p^2 \delta_{ij}+(2\epsilon +\delta^2)p_mp_s]\right]\right. \\ 
&\quad -q_mp_s\left[\delta_{nr}- \gamma \delta \left(p\delta_{nr}+\frac{p_np_r}{p}\right) + \gamma ^2[\epsilon p^2 \delta_{nr}+(2\epsilon +\delta^2)p_np_r]\right]\Bigg\}
\\
&= i \hbar (\epsilon_{mni}\epsilon_{mjr}q_rp_n-\epsilon_{nim}\epsilon_{nsj}q_mp_s)(1-\delta \gamma p+ \epsilon \gamma^2p^2)
\\
&= i\hbar \epsilon_{ijk}L_k (1-\delta \gamma p+ \epsilon \gamma^2p^2), 
\end{split}
\end{equation}
this results in the usual commutator bracket being modified with additional translational momentum terms. We also observe that angular momentum is now not only the generator of rotation but also of translation. As the GUP parameters $\delta = \gamma = \epsilon = 0$, we recover the original commutator algebra. The effects of GUP directly influence the angular momentum term by adding extra terms. From standard quantum mechanics, we can assign $L_{0,k}$ for the orbital angular momentum, following Eq. (\ref{5}). We can expand the modified angular momentum in terms of the generators of rotation and translation, respectively as
\begin{equation}
L_k=L_{0,k}(1-\delta \gamma p+ \epsilon \gamma^2p^2).
\end{equation}
Now, by using GUP deform commutator relation Eq.(\ref{deform}), we write Eq.(\ref{e2}) as:
\begin{equation}
[L_i,[L_j,L_k]]+[L_j,[L_k,L_i]]+[L_k,[L_i,L_j]]=i \hbar \{\epsilon_{ijk}[L_i,L_i]+\epsilon_{ijk}[L_j,L_j]+\epsilon_{ijk}[L_k,L_k]\}(1-\delta \gamma p+ \epsilon \gamma^2p^2)=0,
\end{equation}
it means in the presence of GUP the Jacobi relation remains same.
Furthermore, the results of Eq. (\ref{e3}) also remain the same, allowing us to define the eigenstates of $L^2$ and $L_z$ simultaneously when we consider the GUP modifications. This can be expressed as 
\begin{equation}
[L^2,L_j]= [L_i^2,L_j]+[L_j^2,L_j]+[L_k^2,L_j] =0.
\end{equation}
If we consider the commutator relation between the components of angular momentum and linear momentum, we obtain
\begin{equation}
\begin{split}[L_i,p_m]&=\epsilon_{ijk}[q_j p_k,p_m]= \epsilon_{ijk}[q_j,p_m]p_k\\
&=\epsilon_{ijk} p_k i\hbar\Bigg\{\delta_{jm}- \gamma \delta \left(p\delta_{jm}+\frac{p_jp_m}{p}\right) + \gamma ^2[\epsilon p^2 \delta_{jm}+(2\epsilon +\delta^2)p_jp_m]\Bigg\}\\
&= i \hbar\epsilon_{imk}p_k(1-\gamma \delta p +\epsilon \gamma^2 p^2),
\end{split}
\end{equation}
we observe that the relation is modified by an extra term involving \(\vec{p}\). The original results remain unchanged if we neglect the quantum gravitational parameters.
From teh above results we can easily show that
\begin{equation}\label{e4}
\begin{split}[L_i,p^2] &= [L_i,p_i^2+p_j^2+p_k^2]=p_i[L_j,p_i]+[L_j,p_i]p_i+p_j[L_k,p_j]+[L_k,p_j]p_j \\
&=i \hbar (1-\gamma \delta p +\epsilon \gamma^2 p^2) [\epsilon_{jim}p_ip_m+\epsilon_{jim}p_mp_i+\epsilon_{kjn}p_jp_n+\epsilon_{kjn}p_np_j]
\\
&= 0, 
\end{split}
\end{equation}
by the property of \(\epsilon_{ijk}p_jp_k + \epsilon_{ijk}p_kp_j = 0\),
and utilizing Eq.(\ref{deform}), we notice that from Eq.(\ref{e5}) $L_z$ commutes with $p$, Eq.(\ref{e6}), and Eq.(\ref{e7}) remain unchanged from the standard theory. Consequently, we can define the simultaneous eigenstates for \(L^2\), \(L_z\), and \(p\). Despite the GUP-deformed commutator algebra, all the results remain consistent with the standard relations of quantum mechanics, indicating that the fundamental structure is unchanged.
\subsection{Ladder operators in the presence of GUP}
In this subsection, we calculate the effects of the GUP-deformed algebra on the ladder operators. Using the standard procedure, it will follow the same structure as in the standard theory, but the quantum numbers $l$ and $m$ are now not integers and include correction terms. As mentioned above, we can consider the simultaneous eigenstates of $p$, $L_z$, and $L^2$ as
\begin{equation*}
L^2|p\lambda m \rangle = \lambda \hbar^2 |p\lambda m \rangle,  \quad  L_z|p\lambda m \rangle =m \hbar |p\lambda m \rangle
\end{equation*}
Using the Eq.(\ref{use}), the component of angular momentum and ladder operator does not commute and gives
\begin{equation}\label{pm}
[L_z,L_{\pm}]=[L_z,L_x] \pm i [L_z,L_y]= \hbar(i L_y \pm L_x)(1-\gamma \delta p +\epsilon \gamma^2 p^2 )= \pm \hbar L_{\pm}(1- \zeta),
\end{equation}
where $\zeta$ is the GUP corrected factor, $\zeta = \delta \gamma p-\epsilon \gamma^2 p^2$. The ladder operator commutes with the magnitude of angular momentum as
\begin{equation}\label{mp}
[L^2,L_{\pm}]= [L^2,L_x] \pm i [L^2, L_y] =0
\end{equation}
and this structure is similar to the standard theory as mentioned in Eq.(\ref{ls}).
With the help of GUP-modified commutator relation  Eq.(\ref{use}), we get
\begin{equation}[L_+,L_-] = [L_x+iL_y,L_x-iL_y]= -2i[L_x,L_y]=2 \hbar L_z(1-\zeta).
\end{equation}
It also satisfies,
\begin{equation}\label{l1}
L_{\pm}L_{\mp} = L^2-L_z^2 \pm \hbar L_z(1-\zeta).
\end{equation}
Now, using Eq.(\ref{pm}), Eq.(\ref{4}) became:
\begin{equation}\label{4}
L_zL_{\pm}  |p\lambda m\rangle =L_{\pm}[L_z\pm (1-\zeta)]|p\lambda m\rangle = \hbar[m\pm(1-\zeta)]L_{\pm}|p\lambda m\rangle
\end{equation}
 and using Eq.(\ref{mp}), we get
 \begin{equation}
L^2L_{\pm}|p\lambda m\rangle=L_{\pm}L^2|p\lambda m\rangle= \lambda \hbar^2L_{\pm}|p\lambda m\rangle,
 \end{equation}
and the norm of the state $L_{\pm}|p\lambda m \rangle$ is obtained as
\begin{equation}
||L_{\pm}|p \lambda m \rangle||^2 =\langle p \lambda m| L_{\mp}L_{\pm} |p \lambda m \rangle = \hbar^2[\lambda -m\{m \pm (1-\gamma \delta p +\epsilon \gamma^2 p^2)\}] \geq 0.
\end{equation}
If we consider $L^2$ and $L_z$ only, we obtain
\begin{equation}
||L_{\pm}| \lambda m \rangle||^2 =\langle  \lambda m| L_{\mp}L_{\pm} | \lambda m \rangle = \hbar^2[\lambda -m\{m \pm (1-\langle \zeta \rangle)\}] \geq 0, \label{average}
\end{equation}
the GUP modification appears as $\langle \zeta \rangle$, which is the expectation value of the corrections. The term $1 - \langle \zeta \rangle$ is a positive quantity. We also note from the eigenvalues of $L^2$ and $L_z$ in Eq. (\ref{l1}) that $\langle L^2 \rangle \geq \langle L_z^2 \rangle$, which requires $\lambda \geq m^2$. This implies that there is an upper and lower bound on $m$, denoted as $m_+$ and $m_-$, respectively. Additionally, from Eq. (\ref{average}), we obtain
\begin{equation}
L_{\pm}| p \lambda m_{\pm} \rangle = 0 \implies \lambda = m_{\pm}[m_{\pm}\pm(1- \zeta)] \label{lambda}
\end{equation}
we can get $m_{\pm}$ from any starting value of $m$  by multiplying $u$ and $v$ to $L_+$ and $L_-$ respectively where $u, v \in \mathbb{N}$ as 
\begin{equation}
m_+ = m + u(1- \zeta), \quad  m_- = m - v(1- \zeta)
\end{equation}
and on combining these two relations we get,
\begin{equation}
m_+ = m_- + (u+v)(1- \zeta) = m_- + n(1-\zeta)
\end{equation}
where $n\in \mathbb{N}$. On taking the above equation and  Eq.(\ref{lambda}), one can obtain
\begin{equation}
m_+ = \frac{n}{2}(1- \zeta) = l(1-\zeta), \quad m_- = - \frac{n}{2}(1- \zeta)= -l(1-\zeta),
\end{equation}
and also we get 
\begin{equation}
\lambda = l(l+1)(1-\zeta)^2.
\end{equation}
 We noticed that usually, the magnetic quantum number $m$ might not be an integer because the difference between two consecutive eigenvalues of $L_z$, $m_1$ and $m_2$ the difference is $(1-\zeta)$. Thus, we redefined it as 
 \begin{equation}
    m \to m(1-\zeta)
 \end{equation}
Now, the new GUP-modified magnetic quantum number $m$ has integer upper and lower bounds as 
 \begin{equation}
    -l\leq m \leq l.
 \end{equation}
Next, the eigenvalues of $L^2$ and $L_z$ for any given state are
  \begin{equation}
L^2| p l m\rangle = \hbar^2 l(l+1) (1- \zeta)^2 | p l m \rangle , \quad L_z| p l m \rangle = \hbar (1- \zeta) | p l m \rangle. \label{2eq}
\end{equation}
To show that the eigenvalues of \(L^2\) and \(L_z\) are compatible with the uncertainty relation in Eq. (\ref{use}), we first note that the expectation values of \(L_x\) and \(L_y\) in an eigenstate of \(L^2\) and \(L_z\) are zero:
\begin{equation}
    \langle L_x \rangle = \langle L_y \rangle = 0.
\end{equation}
The variances of these components are given by:
\begin{equation}
    (\Delta L_x)^2 = \langle L_x^2 \rangle, \quad (\Delta L_y)^2 = \langle L_y^2 \rangle.
\end{equation}
Now, for the right-hand side of Eq. (\ref{2eq}), the uncertainty relation between \(L_x\) and \(L_y\) is:
\begin{equation}
    \Delta L_x \Delta L_y \geq \frac{|\langle [L_x, L_y] \rangle|}{2} = \frac{\hbar}{2} |\langle L_z \rangle (1 - \langle \zeta \rangle)| = \frac{\hbar^2}{2} |m| (1 - \langle \zeta \rangle)^2, \label{qq}
\end{equation}
where we used the commutator \([L_x, L_y] = i \hbar L_z\) and the GUP correction factor \((1 - \langle \zeta \rangle)\).
By the equivalence between \(L_x\) and \(L_y\) in an eigenstate of \(L_z\), we insert Eq. (\ref{qq}) into the left-hand side of Eq. (\ref{2eq}), yielding:
\begin{equation}
    \langle L^2 \rangle = \hbar^2 l(l+1)(1-\langle \zeta \rangle)^2 = \langle L_x^2 \rangle + \langle L_y^2 \rangle + \langle L_z^2 \rangle \geq \hbar^2 (m^2 + |m|) (1-\langle \zeta \rangle)^2,
\end{equation}
where the equality holds when \(m = \pm l\). We observe that \(\langle L_z^2 \rangle\) is bounded by \(\langle L^2 \rangle\) due to the uncertainty in the other components, similar to standard quantum mechanics. Now, we will apply the results of the modified angular momentum and ladder operators to a physical system and examine how GUP modifications affect those systems \cite{81,82,83,84,85}.

 \section{Modified Hydrogen atom}\label{sec4}
From the sections above, we now know that by introducing a minimal angular resolution, the angular momentum algebra has changed significantly. To observe these effects, we can apply these corrections to different quantum systems. As we also know, angular momentum algebra plays an important role in understanding atomic systems. In this section, we study the modified energy levels and spectrum of hydrogen atoms based on the GUP-modified angular momentum algebra. We incorporate the effects of GUP directly into the Hamiltonian of the system.

The usual Hamiltonian for hydrogen atom for a central force with the potential energy $U(r)=\frac{e^2}{r}$ is given as
 \begin{equation}
\mathcal{H} \psi(\Vec{r})= \left[\frac{p_r^2}{2m}+\frac{L^2}{2mr^2}-\frac{e^2}{r}\right]\psi(\Vec{r}),
 \end{equation}
we have the radial momentum term $p_r$ and the square of angular momentum which consists of both radial and angular parts. However, we include the GUP effects in terms of expected values \textit{i.e.} $\langle \zeta \rangle = \delta \gamma \langle p \rangle - \epsilon \gamma \langle p^2 \rangle$ as 
 \begin{equation}\label{ran}
  L^2|lm\rangle =\hbar^2 l(l+1)(1-\langle \zeta \rangle)^2|lm\rangle ,          \quad  L_z|lm\rangle = \hbar m(1-\langle \zeta \rangle)|lm\rangle, \quad  \Vec{p} |lm\rangle=\Vec{p_0}(1-\langle \zeta \rangle)|lm\rangle.
 \end{equation}
In this context, \(\langle \zeta \rangle\) represents the average value of the GUP corrections. It is crucial to assume expectation values for observable physical quantities when considering Planck-scale corrections. From Eq. (\ref{ran}), we observe that \(\vec{L}\) is proportional to \(\vec{L}_0\), and both \(L_i\) and \(L^2\) commute with the Hamiltonian \(\mathcal{H}\). Using this result, the GUP-modified radial part of the Schr\"odinger
 equation becomes
 \begin{equation}
\left[\frac{p_{0,r}^2}{2m}(1-\langle \zeta \rangle)^2+\frac{\hbar^2l(l+1)}{2mr^2}(1-\langle \zeta \rangle)^2-\frac{e^2}{r}\right]y_l(r)=Ey_l(r). \label{hydrogen}
 \end{equation}
where \( E \) is the energy eigenvalue. The above Eq. (\ref{hydrogen}) consists of two parts one is the unperturbed part without the GUP modification, expressed in terms of the unmodified momentum and angular momentum, and other is the perturbative part that includes the corrections. It can be written as:
\begin{equation}
  \left[\frac{p_{0,r}^2}{2m} + \frac{\hbar^2 l(l+1)}{2mr^2} - \frac{e^2}{r}\right] y_l(r) + \left[\frac{p_{0,r}^2}{2m} + \frac{\hbar^2 l(l+1)}{2mr^2}\right] \{\langle \zeta \rangle^2 - 2\langle \zeta \rangle\} y_l(r) = E y_l(r).
\end{equation}
Here, the first term represents the low-energy part of the Hamiltonian without GUP corrections. Now on choosing the variables like,
\begin{equation}
\chi = \sqrt{\frac{-2mE}{\hbar^2}}, \quad a = \frac{\hbar^2}{me^2}, \quad \nu = \frac{1}{a\chi}, \quad z = \frac{2\chi r}{1 - \langle \zeta \rangle}, \quad y_l(r) = z^{(l+1)} e^{-\frac{z}{2}} v(z)
\end{equation}
and substituting these variables in Eq.(\ref{hydrogen}) gives
\begin{equation}\label{deq}
\left[z\frac{d^2}{dz^2} +(2l+2-z)\frac{d}{dz}+\left(\frac{\nu}{1-\langle \zeta \rangle} -l -1\right)\right]=0.
\end{equation}
On taking $\gamma=0$ the above equation changes to the well-known Laguerre equation.
With the given condition as 
\begin{equation}
n'=\frac{\nu}{1-\langle \zeta \rangle} -l -1 \in \mathbb{N}.
\end{equation}
On solving the Eq.(\ref{deq}) we obtain the associated Laguerre polynomial
\begin{equation}
L^{(2l+1)}_{n'}(z) = \sum_{i=0}^{n'} \frac{(-1)^i (n'+2l+1)! \, z^i}{(n'-i)! \, (2l+1+i)! \, i!}.
\end{equation}
then, the solution for the radial Schrodinger equation (\ref{hydrogen}) is 
\begin{equation}
 y_l(r)= z^{(l+1)} e^{-\frac{z}{2}} \sum_{i=0}^{n'} \frac{(-1)^i (n'+2l+1)! \, z^i}{(n'-i)! \, (2l+1+i)! \, i!},   
\end{equation}
and the generalized principal quantum number is 
\begin{equation}
n=\nu=\frac{e^2}{\hbar}\sqrt{\frac{m}{-2E}}= (n'+l+1)(1-\langle \zeta \rangle)= n_0(1-\langle \zeta \rangle),
\end{equation}
where $n_0$ is the principal quantum number. The GUP-deformed expression of the energy level of the hydrogen atom is
\begin{equation}
    \begin{split}
 E_n &=-\frac{2\pi e^4}{h^2} \frac{m}{[(n'+l+1)(1-\langle \zeta \rangle)]^2} \\
    &=-\frac{e^4}{h^2c^2} \frac{mc^2}{n^2} \\
    &\approx E^{(0)}_n[1+2\delta \gamma \langle p_0 \rangle +\gamma^2(3 \delta^2 \langle p_0 \rangle ^2 -2 \epsilon \langle p_0^2 \rangle)],
\end{split}
\end{equation}
$E^{(0)}_n$ represents the energy for the corresponding energy level in the standard theory. With GUP corrections, the energy for the hydrogen atom now includes additional terms involving the expectation values of $p_0$ and $p_0^2$ in a $|n_0 l m \rangle$ eigenstate. If we consider no GUP effects by setting $\gamma = 0$, we recover the original energy level expression. In standard QM, the energy emitted or absorbed during a transition between two energy levels is given by the energy difference between the two states. Thus, it is straightforward to determine the frequency and wavelength of the emitted or absorbed lines through the wavelength and differrence in energy level relation. It is also well known that the energy of the hydrogen atom is inversely proportional to the square of the \textit{principal quantum number}. From this, we can also calculate the wavelength of the emitted photon when the atom transitions from energy level $E_i$ to $E_f$ as:

\begin{equation}\label{r}
\frac{1}{\lambda} = \frac{|E_i - E_f|}{hc} = R_\infty \left| \frac{1}{n_{0,f}^2 (1 - \langle \zeta_f \rangle)^2} - \frac{1}{n_{0,i}^2 (1 - \langle \zeta_i \rangle)^2} \right|,
\end{equation}

where $R_{\infty}$ is the Rydberg constant. Equation (\ref{r}) shows that the GUP-modified spectrum not only depends on the \textit{principal quantum number} but also includes corrections involving the expectation values of the electron's momentum, as well as its angular momentum quantum numbers.

\section{Conclusion}

Quantum mechanics, as is well known, is based on the Heisenberg uncertainty relation. Many quantum gravity theories predict the need for modifications that incorporate the concept of a minimal length. With the development of various quantum gravity theories, identifying a distinct signature of quantum gravity remains a significant challenge, making experimental testing crucial for measuring Planck scale effects in low-energy quantum systems.

In this review, we study the effects of the GUP on angular momentum algebra and how it modifies systems where angular momentum is required. We begin by reviewing the origin of the GUP and the higher-order GUP introduced by Nouicer and Pedram in Sec. \ref{sec1}. We also discuss the deformed commutator bracket proposed by Ali, Das, and Vagenas, along with the derivation of the deformed canonical variables.

In this review, we consider only the deformed commutator bracket containing linear and quadratic momentum terms and proceed to explore its implications for angular momentum algebra. Our observations reveal that the GUP-modified algebra between position and momentum introduces modifications to the angular momentum commutator bracket, adding a term that accounts for the GUP modification. The comparison between the standard angular momentum algebra and the GUP-modified angular momentum algebra is summarized in Table \ref{tab}.
\renewcommand{\arraystretch}{1.5} 
\begin{table}[ht]
    \centering
    \caption{Comparison of standard Angular Momentum Algebra and GUP-Modified Algebra}
    \begin{tabular}{@{}l l l@{}}
        \toprule
        \textbf{Aspect}                                   & \textbf{Standard Angular Momentum}                      & \textbf{GUP-Modified Algebra}                          \\ \midrule
        Commutator Relations                 & $[L_i, L_j] = i\hbar \epsilon_{ijk} L_k$    & $[L_i, L_j] = i\hbar \epsilon_{ijk} L_k(1 - \delta \gamma p + \epsilon \gamma^2 p^2)$ \\
        Jacobi Identity                      &  $[L_i, [L_j, L_k]] + [L_j, [L_k, L_i]] + [L_k, [L_i, L_j]] = 0$ & $[L_i, [L_j, L_k]] + [L_j, [L_k, L_i]] + [L_k, [L_i, L_j]] = 0$                 \\
        Commutators with $L^2$            & $[L^2, L_j] = 0$                            & $[L^2, L_j] = 0$                                     \\
        Commutators with Momentum            & $[L_i, p_j] = i\hbar \epsilon_{ijk} p_k$   & $[L_i, p_j] = i\hbar \epsilon_{ijk} p_k (1 - \gamma \delta p + \epsilon \gamma^2 p^2)$ \\
        Commutators with $p^2$            & $[L_i, p^2] = 0$                           & $[L_i, p^2] = 0$                                     \\
        Eigenstates of $L^2$ and $L_z$  & Can be simultaneously defined                 & Can still be defined simultaneously.                  \\
        Commutator with $L_z$              & $[L_z, L_{\pm}] = \pm \hbar L_{\pm}$       & $[L_z, L_{\pm}] = \pm \hbar L_{\pm}(1 - \zeta)$     \\
        Norm of the State                    & $||L_{\pm} |l, m\rangle ||^2 \geq 0$        & $||L_{\pm} |p, \lambda, m\rangle ||^2 \geq 0$ with GUP corrections. \\
        Quantum Numbers      & Integer values for $l$ and $m$            & Integer values but with bounds on $m$ due to GUP.    \\
       \bottomrule
    \end{tabular}\label{tab}
\end{table}

In Sec. \ref{sec4}, we carry out the application of these modifications and examine how they affect the energy levels and energy spectrum of the hydrogen atom. We observe that the principal quantum number changes due to the modification, with additional terms representing the average values of \( p \) and \( p^2 \), which affect the energy levels of the hydrogen atom. We also note that the energy spectrum now includes corrections that account for the expected values of the electron's momentum and angular momentum quantum number.

Since higher-order GUPs will also follow the same algebra with additional higher-order terms, we expect the algebra to remain consistent for these higher-order GUPs also. With these modifications to the angular momentum algebra, new avenues for phenomenological testing and theories are opened, which can potentially be tested in the future in contexts such as the Zeeman effect, Stark effect, simple harmonic oscillator (SHO), and others. This modification to the angular momentum algebra provides an opportunity to tighten the upper bound of the GUP parameter as studied in \cite{Sau} by considering systems with angular momentum like Hydrogen atom in our case. We leave this for our future endeavors.




\begin{thebibliography}{99}
\bibitem{rana1987investigation} N. C. Rana, ``An investigation of the motions of the node and perihelion of Mercury", 
\emph{A \& A}, \textbf{181}, 195(1987)
\bibitem{genov2009mimicking} Genov \textit{et al.}, ``Mimicking celestial mechanics in metamaterials", 
\emph{Nat. Phys.}, \textbf{5}, 687(2009)
\bibitem{wojtak2011gravitational} R. Wojtak, S. H. Hansen, J. Hjorth, ``Gravitational redshift of galaxies in clusters as predicted by general relativity", 
\emph{Nature}, \textbf{477}, 567(2011).
 \bibitem{cite1} J. D. Bekenstein, ``Black holes and entropy", 
 \emph{Phys. Rev. D},
 \textbf{7}, 2333 (1973).
 \bibitem{cite2} S. W. Hawking, ``Black hole explosions?", 
 \emph{Nature},
 \textbf{248}, 30 (1974).
\bibitem{cite3} S. W. Hawking, ``Black holes and thermodynamics", 
\emph{Phys. Rev. D},
\textbf{13}, 191 (1976).
\bibitem{cite4} S. Chandrasekhar and K. S. Thorne, ``The mathematical theory of black holes", 
\emph{Am. J. Phys.},
\textbf{53}, 1013 (1985).
\bibitem{cite5} J. Kormendy and D. Richstone, ``Inward Bound---The Search For Supermassive Black Holes In Galactic Nuclei", \emph{Ann. Rev. Astron. Astrophys.},
\textbf{33(1)},
 581 (1995).

 \bibitem{63}S. W. Hawking, ``Particle Creation by Black Holes", \emph{Commun. Math. Phys.}, \textbf{43}, 199 (1975) [Erratum-ibid. 46, 206 (1976)]. 
 \bibitem{64} W. G. Unruh, ``Sonic analog of black holes and the effects of high frequencies on black hole evaporation", \emph{Phys. Rev. D}, \textbf{51}, 2827 (1995).
 
 \bibitem{guth1982fluctuations} A.H. Guth, S.Y. Pi,   ``Fluctuations in the new inflationary universe", 
\emph{ Phys. Rev. Lett.}, \textbf{49}, 1110(1982)
\bibitem{linde1982new} A.D. Linde,   ``A new inflationary universe scenario: a possible solution of the horizon, flatness, homogeneity, isotropy and primordial monopole problems", 
\emph{ Phys. Rev. B}, \textbf{108}, 389(1982)
\bibitem{bardeen1983spontaneous} J.M. Bardeen, P.J. Steinhardt, M.S. Turner,   ``Spontaneous creation of almost scale-free density perturbations in an inflationary universe", 
\emph{ Phys. Rev. D}, \textbf{28}, 679(1983)
\bibitem{ashtekar2005quantum} A. Ashtekar and M. Bojowald,   ``Quantum geometry and the Schwarzschild singularity", 
\emph{Classical and Quantum Gravity}, \textbf{23}, 391(2005)

\bibitem{sing1},
 ``Perturbations in tachyon dark energy and their effect on matter clustering",
  \emph{J. Cosmol. Astropart. Phys.}, \textbf{2020}, 8(2020)

\bibitem{ashtekar2009loop} A. Ashtekar,   ``Loop quantum cosmology: an overview", 
\emph{Gen. Relativ. Gravit.}, \textbf{41}, 707(2009).
\bibitem{veneziano1986stringy} G. Veneziano,   ``A stringy nature needs just two constants", 
\emph{Europhys. Lett.}, \textbf{2}, 199(1986)

\bibitem{witten1996reflections} E. Witten,   ``Reflections on the fate of spacetime", 
\emph{Phys. Today}, \textbf{49}, 24(1996)

\bibitem{scardigli1999generalized} F. Scardigli,   ``Generalized uncertainty principle in quantum gravity from micro-black hole gedanken experiment", 
\emph{Phys. Lett. B}, \textbf{452}, 39(1999)

\bibitem{gross1988string} D. J. Gross, P. F. Mende,   ``String theory beyond the Planck scale", 
\emph{Nucl. Phys. B.}, \textbf{303}, 407(1988)

\bibitem{amati1989can} D. Amati, M. Ciafaloni, G. Veneziano,   ``Can spacetime be probed below the string size?", 
\emph{Phys. Lett. B}, \textbf{216}, 41(1989)

\bibitem{yoneya1989interpretation} T. Yoneya,   ``On the interpretation of minimal of minimal length in string theories", 
\emph{Gen. Relativ. Gravit.}, \textbf{4}, 16(1989)
\bibitem{s3} C. A. Mead,   ``String theory, supersymmetry, unification, and all that", 
\emph{Rev. Mod. Phys.}, \textbf{71}, S112(1999).
\bibitem{sharma2019background} M. Sharma, T. Zhu and A Wang,   ``Background dynamics of pre-inflationary scenario in Brans-Dicke loop quantum cosmology", 
\emph{Commun. Theor. Phys.}, \textbf{71}, 1205(2019)

\bibitem{garay1995quantum} L.J. Garay,   ``Quantum gravity and minimum length", 
\emph{Int. J. Mod. Phys.}, \textbf{10}, 145(1995)
\bibitem{as1} A. Ashtekar, S. Fairhurst and J.L. Willis,   ``Quantum gravity, shadow states and quantum mechanics", 
\emph{Classical and Quantum Gravity}, \textbf{20}, 1031(2003)

\bibitem{ha1} G.M. Hossain, V. Husain and S.S. Seahra,   ``Background-independent quantization and the uncertainty principle", 
\emph{Gen. Relativ. Gravit.}, \textbf{27}, 165013(2010)
\bibitem{rovelli1998strings} C. Rovelli,   ``Strings, loops and others: a critical survey of the present approaches to quantum gravity", 
\emph{arXiv preprint gr-qc/9803024}, (1998)
\bibitem{sh1} M. Sharma, T. Zhu, A. Wang,
``Background dynamics of pre-inflationary scenario in Brans-Dicke loop quantum cosmology",
 \emph{Commun. Theor. Phys.}, \textbf{71}, 1205(2019).

 

\bibitem{68}D. J. Gross, P. F. Mende, ``String Theory Beyond the Planck Scale", \emph{Nucl. Phys. B}, \textbf{303}, 407 (1988). 

\bibitem{69}D. Amati, M. Ciafaloni and G. Veneziano, ``Classical and Quantum Gravity Effects from Planckian Energy Superstring Collisions", \emph{Int. J. Mod. Phys. A}, \textbf{3}, 1615 (1988).

\bibitem{70}D. Amati, M. Ciafaloni and G. Veneziano, ``Superstring Collisions at Planckian Energies", \emph{Phys. Lett. B}, \textbf{197}, 81 (1987).

\bibitem{71}D. Amati, M. Ciafaloni and G. Veneziano,``Higher Order Gravitational Deflection And Soft Bremsstrahlung In Planckian Energy Superstring Collisions", \emph{Nucl. Phys. B}, \textbf{347}, 550 (1990). 

\bibitem{72}G. M. Hossain, V. Husain and S. S. Seahra, ``Background independent quantization and the uncertainty principle", \emph{Class. Quant. Grav.}, \textbf{27}, 165013 (2010) arXiv:1003.2207[gr-qc]. 
 \bibitem{double1} Giovanni Amelino-Camelia,
 ``Doubly-special relativity: Facts, myths and some key open issues",
\emph{Symmetry},\textbf{2}, 230(2010).

\bibitem{gh1} S. Ghosh,
  ``Lagrangian for doubly special relativity particle and the role of noncommutativity",
 \emph{Phys. Rev. D}, \textbf{74},084019(2006).

 \bibitem{54} A. Tawfik and A. Diab, ``Review on Generalized Uncertainty Principle", \emph{Reports on Progress in Physics}, \textbf{78}, 12 (2015). 

\bibitem{55} S. Hossenfelder, ``Minimal length scale scenarios for quantum gravity", \emph{Living Rev. Rel.}, \textbf{16}, 2 (2013)1203.6191[gr-qc].

\bibitem{56} A. Peres and N. Rosen, ``Quantum Limitations on the Measurement of Gravitational Fields", \emph{Phys. Rev.}, \textbf{118}, 335 (1960). 
 
  
\bibitem{57} H. S. Snyder, ``Quantized space-time", \emph{Phys. Rev.}, \textbf{71}, 38 (1947).

\bibitem{58} C. N. Yang, ``On quantized space-time", \emph{Phys. Rev.}, \textbf{72}, 874 (1947). 

\bibitem{59} E. P. Wigner, ``Relativistic Invariance and Quantum Phenomena", \emph{Rev. Mod. Phys.}, \textbf{29}, 255 (1957). 

\bibitem{60} H. Salecker and E. P. Wigner, ``Quantum limitations of the measurement of space-time distances", \emph{Phys. Rev.}, \textbf{109}, 571 (1958).

\bibitem{61} C. A. Mead, ``Possible Connection Between Gravitation and Fundamental Length",\emph{ Phys. Rev. D} , \textbf{135}, B849 (1964).

\bibitem{62}C. A. Mead, ``Observable Consequences of Fundamental-Length Hypotheses", \emph{Phys. Rev.}, \textbf{143}, 990 (1966). 

\bibitem{65}S. Majid and H. Ruegg, ``Bicrossproduct structure of kappa Poincare group and noncommutative geometry", \emph{Phys. Lett. B},\textbf{ 334}, 348 (1994) [hep-th/9405107].

\bibitem{66}A. Kempf, ``Uncertainty relation in quantum mechanics with quantum group symmetry", \emph{J. Math. Phys.}, \textbf{35}, 4483 (1994) hep-th/9311147. 

\bibitem{67} A. Kempf, ``Quantum field theory with nonzero minimal uncertainties in positions and momenta", \emph{Preprint DAMTP/94-33}, (1994) [hep-th/9405067].

\bibitem{100} G. Bhandari \textit{et al.}, ``Quantum deformed phantom dynamics in light of the generalized uncertainty principle",
\emph{General Relativity and Gravitation},
 \textbf{56},
139 (2024),.

\bibitem{101} G. Bhandari \textit{et al.}, ``Generalized uncertainty principle distorted quintessence dynamics",
 \emph{arXiv preprint arXiv:2405.08680},
2024.

\bibitem{102} G. Bhandari \textit{et al.}, ``Quantum Gravity Corrections to Hawking Radiation via GUP",
 \emph{arXiv preprint arXiv:2407.19268},
2024.
\bibitem{30} S.Das and E.C.Vagenas, ``Phenomenological implications of the generalized uncertainty principle", \emph{Canadian Journal of Physics}, \textbf{87}, 233(2009). 


\bibitem{31} A.F.Ali, S.Das, and E.C.Vagenas, ``A proposal for testing quantum gravity in the lab", \emph{Phys.Rev.D}, \textbf{84}, 44013(2011). 

\bibitem{32} A.F.Ali, S.Das, and E.C.Vagenas, ``Discreteness of space from the generalized uncertainty principle", \emph{Phys. Lett. B}, \textbf{678}, 497(2009). 

\bibitem{32a} S. Deb, S. Das, E. C. Vagenas, ``Discreteness of space from GUP in a weak gravitational field",
  \emph{Physics Letters B},
 \textbf{755},
 17 (2016).
\bibitem{33} I.Pikovski \textit{et al.}, ``Probing Planck-scale physics with quantum optics", Nature Physics, \textbf{8}, 393(2012).
\bibitem{G8} G. Bhandari \textit{et al.},
 ``Generalized Uncertainty Principle and the Zeeman Effect: Relativistic Corrections Unveiled",
 \emph{arXiv:2410.11965}, 2024.
\bibitem{34} M. Maggiore, ``A generalized uncertainty principle in quantum gravity", \emph{Phys. Lett. B}, \textbf{304}, 65(1993).
\bibitem{35} F.Scardigli, 
 ``Generalized uncertainty principle 
 in quantum gravity from micro-black hole gedanken experiment", \emph{Phys. Lett. B}, \textbf{452}, 
 39(1999).
\bibitem{36} A. Kempf, G. Mangano, and R.B.Mann, ``Hilbert space representation of the minimal length uncertainty 
 relation", \emph{Phys. Rev. D}, \textbf{52}, 1108(1995).
\bibitem{37} E. Witten, ``String theory dynamics in various dimensions", \emph{Nuclear Physics B},\textbf{443}, 1(1995).
\bibitem{38} C. Bachas, ``The short distance structure of open string theory" arXiv preprint hep-th/9907023 (1999).
\bibitem{Z2} A. Kempf,``Non-point like particles in harmonic oscillators",  \emph{J. Phys. A: Math. Gen.}, \textbf{30}, 2093 (1997).

\bibitem{Z3} A. Kempf, G. Mangano, R.B. Mann, ``Hilbert space representation of the minimal length uncertainty relation" \emph{Phys. Rev. D}, \textbf{52}, 1108 (1995).

 \bibitem{Z4} A. Kempf, G. Mangano, ``Minimal length uncertainty relation and ultraviolet regularization", \emph{Phys. Rev. D}, \textbf{55}, 7909 (1997).
\bibitem{42} K. Nouicer,``Quantum-corrected black hole thermodynamics to all orders in the Planck length", \emph{Phys. Lett. B} \textbf{646}, 63 (2007).
\bibitem{43} P. Pedram, ``A Higher Order GUP with Minimal Length Uncertainty and Maximal Momentum ",\emph{Phys. Lett. B} 714, 317 (2012).
\bibitem{44} A. F. Ali, S. Das, and E. C. Vagenas,
 ``Discreteness of space from the generalized uncertainty principle", \emph{Physics Letters B},
 \textbf{678},
 497 {2009}.
 


\bibitem{73} L. J. Garay, ``Quantum gravity and minimum length", \emph{Int. J. Mod. Phys. A}, \textbf{10}, 145 (1995) [gr-qc/9403008].


\bibitem{74} K. Nozari and B. Fazlpour, ``Generalized uncertainty principle, modified dispersion relations and early universe thermodynamics", \emph{Gen. Relat. Gravit.}, \textbf{38}, 1661 (2006) [gr-qc/ 0601092]. 

\bibitem{75} R. J. Adler and D. I. Santiago, ``On gravity and the uncertainty principle", \emph{Mod. Phys. Lett. A}, \textbf{14}, 1371 (1999) [gr-qc/9904026]. 

\bibitem{76} S. Hossenfelder, ``Interpretation of quantum field theories with a minimal length scale", \emph{Phys. Rev. D}, \textbf{73}, 105013 (2006) [hep-th/0603032].

\bibitem{77}I. Dadic, L. Jonke and S. Meljanac, ``Harmonic oscillator with minimal length uncertainty relations and ladder operators", \emph{Phys. Rev. D}, \textbf{67}, 087701, (2003) [hep-th/0210264]. 

\bibitem{78} C. Quesne and V. M. Tkachuk, ``Dirac oscillator with nonzero minimal uncertainty in position", \emph{J. Phys. A}, \textbf{38}, 1747 (2005) [math-ph/0412052]. 

\bibitem{79} C. Quesne and V. M. Tkachuk, ``Lorentz-covariant deformed algebra with minimal length and application to the 1+1-dimensional Dirac oscillator", \emph{J. Phys. A}, \textbf{39}, 109090 (2006) [quant-ph/0604118]. 
 \bibitem{45}  X.L. Ka., ``Advanced Quantum Mechanics", 1st edn. Higher Education, Beijing (2001), p. 135.
 \bibitem{46} K. Kowalski, J. Rembielinski, ``Quantum mechanics on a sphere and coherent states", \emph{J. Phys. A}, \textbf{33}, 6035 (2000).
 \bibitem{47} Z.X. Ni, ``Nonlinear Lie algebra and ladder operators for orbital angular momentum", \emph{J. Phys. A},\textbf{32}, 2217 (1999).
 \bibitem{48} D. Ruan, W. Ruan, ``Boson realization of nonlinear SO(3) algebra", \emph{Phys. Lett. A}, \textbf{263}, 78 (1999).
\bibitem{49} J. J. Sakurai, ``Modern Quantum Mechanics", Addison-Wesley, New York (1994). 
\bibitem{50} R. Shankar, ``Principles of Quantum Mechanics", 2nd edn. Plenum, New York (1994).
\bibitem{51} A. Farag, S. Das, E.C.Vagenas, ``The generalized uncertainty principle and quantum gravity phenomenology", \emph{In The Twelfth Marcel Grossmann Meeting: On Recent Developments in Theoretical and Experimental General Relativity, Astrophysics and Relativistic Field Theories (In 3 Volumes)}, 2407 (2012).
\bibitem{52} P. Bosso, and Saurya Das. ``Generalized ladder operators for the perturbed harmonic oscillator." \emph{Annals of Physics}, \textbf{396}, 254 (2018).
\bibitem{53} P. Bosso, S. Das, ``Generalized uncertainty principle and angular momentum", \emph{Annals of Physics}, \textbf{383}, 416 (2017).









\bibitem{81} F. Brau, ``Minimal length uncertainty relation and hydrogen atom", \emph{J. Phys. A}, \textbf{32}, 7691 (1999) [quant-ph/9905033]. 

\bibitem{82} S. Benczik, L. N. Chang, D. Minic and T. Takeuchi, ``The Hydrogen atom with minimal length", \emph{Phys. Rev. A}, \textbf{72}, 012104 (2005) [hep-th/0502222]. 

\bibitem{83} M. M. Stetsko and V. M. Tkachuk, ``Perturbation hydrogen-atom spectrum in deformed space with minimal length", \emph{Phys. Rev. A}, \textbf{74}, 012101 (2006) [quant-ph/0603042]. 

\bibitem{84} M. M. Stetsko, ``Corrections to the ns-levels of hydrogen atom in deformed space with minimal length", \emph{Phys. Rev. A}, \textbf{74}, 062105 (2006) [quant-ph/0703269].

\bibitem{85} M. M. Stetsko and V. M. Tkachuk, ``Orbital magnetic moment of the electron in the hydrogen atom in a deformed space with minimal length", \emph{Phys. Lett. A}, \textbf{372}, 5126 (2008) [0710.5088[quant-ph]].
\bibitem{Sau} S. Das, E. C. Vagenas,
 ``Universality of quantum gravity corrections",
 \emph{Physical review letters},
  \textbf{101},
 221301(2008).
\end{thebibliography}
\end{document}